\documentclass[twocolumn]{aa}
\usepackage{graphicx}
\usepackage{txfonts}

\begin{document}

\authorrunning{P. Hennebelle et al.}

\title{A dynamical model for the dusty ring in the Coalsack}

\titlerunning{The ring in the Coalsack}

\author{P. Hennebelle\inst{1}, A. P. Whitworth\inst{2} \and S. P. Goodwin\inst{3}}

\offprints{P. Hennebelle}

\institute{\'Ecole Normale Sup\'erieure, Laboratoire de Radioastronomie Millimetrique, 
24 Rue Lhomond, 75231 Paris Cedex 05, France \\
\email{Patrick.Hennebelle@ens.fr} \and
School of Physics \& Astronomy, Cardiff University, 
5 The Parade, Cardiff CF24 3YB, Wales, UK \\
\email{Anthony.Whitworth@astro.cf.ac.uk} \and
Department of Physics \& Astronomy, University of Sheffield, Hicks Building, 
Housfield Road, Sheffield S3 7RH \\
\email{S.Goodwin@sheffield.ac.uk}}

\date{Received; accepted}

\abstract{
Lada et al. recently presented a detailed near-infrared extinction map 
of Globule G2 in the Coalsack molecular cloud complex, showing that this starless core has a 
well-defined central extinction minimum. We propose a model for G2 in which a rapid increase 
in external pressure is driving an approximately symmetric compression wave into the core. 
The rapid increase in external pressure could arise because the core has recently been 
assimilated by the Coalsack cloud complex, or because the Coalsack has recently been created by 
two large-scale converging flows. The resulting compression wave has not yet converged on the centre 
of the core, so there is a central rarefaction. The compression wave has increased the density 
in the swept-up gas by about a factor of ten, and accelerated it inwards to speeds of order 
$0.4\,{\rm km}\,{\rm s}^{-1}$. It is shown that even small levels of initial turbulence 
  destroy the ring seen in projection almost completely. In the scenario of strong 
external compression that we are proposing this  implies that the initial turbulent 
energy in this globule is such that $E_{{\rm turb}} / E _{{\rm grav}} \le 2 \%$. 
Protostar formation should occur in about $40,000\,{\rm years}$. 
}
\titlerunning{A dynamical model for the dusty ring in the Coalsack}

\maketitle

\section{Introduction}

In the current paradigm for star formation, single stars, binary and multiple stars, and even 
small-N clusters, are presumed to form following the collapse and fragmentation of a dense 
molecular core (e.g. Andr\'e, Ward-Thompson \& Barsony, 2000). However, many details of this paradigm 
remain to be understood, in particular the processes which form dense molecular cores in the 
first place, and thereby determine the initial conditions for collapse and fragmentation.

In an attempt to understand these processes, Lada and colleagues have explored the structures 
of several cores, using the NICE technique (Alves et al. 1998). These include B68 (Alves, Lada 
\& Lada (2001), where there is evidence that the core is oscillating, in accordance with the 
predictions of Matsumoto \& Hanawa (2003) and Hennebelle (2003), and B335 (Harvey et al. 2001), 
where the density profile suggests that the core has already become unstable against collapse. 
In a more recent application of the NICE technique, Lada et al. (2004) have mapped Tapia's 
Globule 2 in the Coalsack (Tapia, 1973).

The Coalsack is an extended cloud complex at a distance $D \sim 150 \,{\rm to}\, 175\,{\rm pc}$ 
(Cambresy, 1999; Rodgers 1960). If we adopt the smaller distance, the angular extent of the 
complex ($\sim 6^\circ$) corresponds to a linear extent of $\sim 15\,{\rm pc}$, and the mass 
of molecular gas is $\sim 3,000\,{\rm M}_{_\odot}$ (Nyman, Bronfman \& Thaddeus, 1989). The 
highly structured molecular gas mapped by Nyman et al. is probably accompanied by an 
envelope and/or infill of somewhat more diffuse HI, providing shielding from the ambient 
H$_2$-dissociating UV radiation field. This HI would be very hard to detect unambiguously, 
since the systemic velocity of the Coalsack is very low ($\sim -\,5\;{\rm to}\;0\;{\rm 
km}\,{\rm s}^{-1}$) and there would therefore be confusion with local $21\,{\rm cm}$ emitting gas. 

Despite its molecular content, and therefore presumably relatively high density, the Coalsack 
cloud complex shows no unambiguous evidence for ongoing star formation. There are no flare 
stars (Weaver 1973); no T Tauri stars (Weaver 1974a,b); no H$\alpha$ emission stars (Schwartz 
1977); and no IRAS sources associated with either the extinction peaks or CO emission peaks 
(Nyman, Bronfman \& Thaddeus, 1989). Reipurth detected a single HH object, R10, in the 
direction of the Coalsack, and Eaton et al. (1990) showed it to be a bipolar nebula with a 
heavily obscured illuminating star. However, Kato et al. (1999) question whether R10 is 
really associated with the Coalsack, and it is certainly a long way from Globule 2 (hereafter G2).

It is therefore possible that G2, which appears to be the densest core in the Coalsack, is 
in an early stage of condensing to form a protostar (or protostars) and, if so, its detailed 
structure could contain important clues about how cores form. To this end, Lada et al. 
(2004) have obtained deep near-IR photometry of $\sim\,24,000$ stars seen through G2, and 
have used this data to construct an extinction map of the core with unprecedented resolution.

The most remarkable feature of this map is that the core has a local extinction minimum in 
the middle, and the maximum extinction is in a ring-like structure around this minimum. Lada 
et al. discuss some possible explanations for this structure, under the general headings: 
`flattened geometry and magnetic fields' and `spherical geometry: a core in transition'. 
We here present a detailed model under the second heading, in which G2 is a core responding 
to a sudden increase in external pressure as investigated in Hennebelle et al. (2003, 2004). 
This increase in external pressure is driving 
a compression wave into the cloud, producing a dense shell which -- seen in projection 
-- is visible as a ring of enhanced extinction. The broad $^{12}$C$^{18}$O ($J = 2 
\rightarrow 1$) emission line observed by Lada et al. along a single line of sight through 
the core is then attributable to the extra trans-sonic velocity generated by the 
inward-propagating compression wave.

In Section 2 we describe briefly the code used to simulate this model, and the initial 
conditions adopted. In Section 3 we present the results obtained and compare them to the 
observations; it should be emphasised at the outset that we do not perform an extensive 
parameter search to obtain the best fit to the observational data, since this does not, at 
the present time, seem justified by the scope of the observations. In Section 4 we discuss 
the results, attempt to justify key aspects of the model, and summarise our main conclusions.

\section{Numerical method and initial conditions}

\begin{figure*}
\label{FIG:COLUMN}
\setlength{\unitlength}{1mm}
\begin{picture}(40,100)
\includegraphics{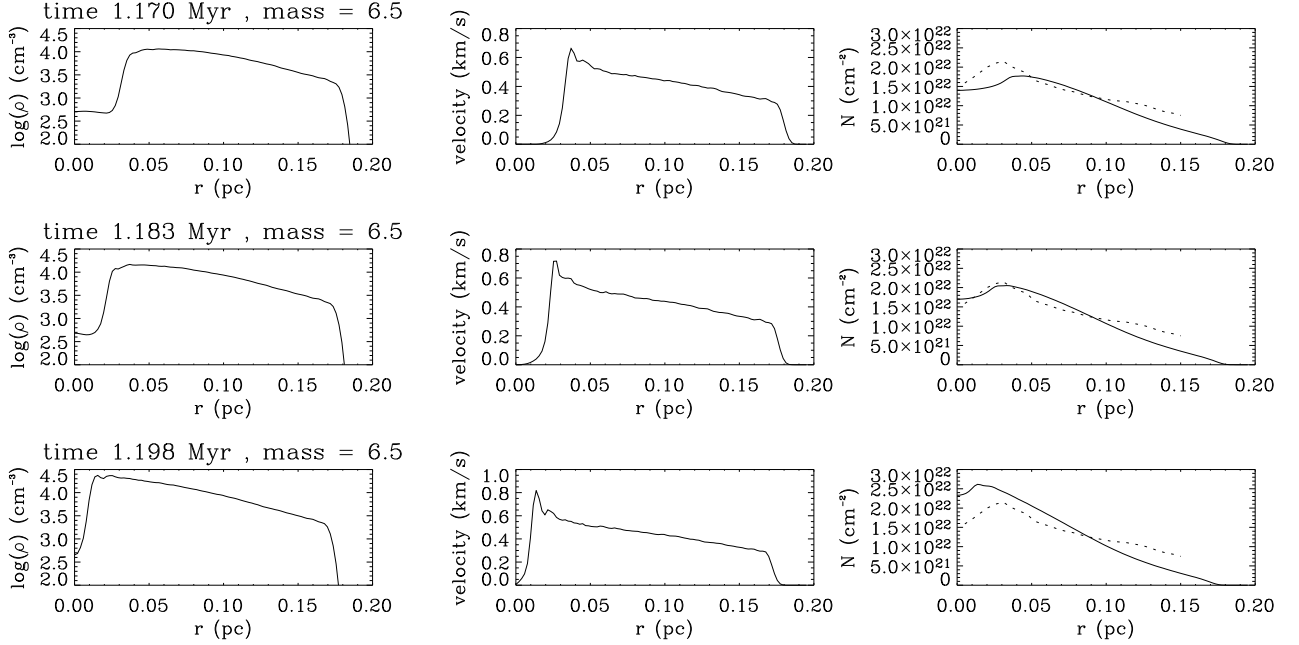}
\end{picture}
\caption{The lefthand column shows volume-density profiles, $n(r)$ (where $n$ is in 
H$_2\,{\rm cm}^{-3}$ and $r$ is in pc. The middle column sows radial velocity profiles, 
$u(r)$ (where $u$ is in ${\rm km}\,{\rm s}^{-1}$) and $r$ is in pc). The righthand 
column shows column-density profiles $N(b)$ (where $N$ is is H$_2\,{\rm cm}^{-2}$ and 
$b$ is in pc. The simulation results are represented by solid lines, and the 
column-density profile derived from the observed extinction is represented by a dashed 
line. The top row is at time $t = 1.170\,{\rm Myr}$; the middle row is at time $t = 
1.183\,{\rm Myr}$; and the bottom row is at time $t = 1.198\,{\rm Myr}$.} 
\end{figure*}

The simulations were performed using the SPH code DRAGON, with $\sim 500,000$ particles. 
This code invokes an oct-tree to find neighbours and calculate gravitational accelerations. 
The particle smoothing lengths are adjusted so that each particle has ${\cal N}_{_{\rm NEIB}} 
= 50 \pm 5$ neighbours. A second-order Runge-Kutta integration scheme is used, with 
multiple particle time-steps. Other details of the code are given in Goodwin, Whitworth \& 
Ward-Thompson (2004).

In the first instance we present a simulation with zero turbulence, since this seems to 
capture well the azimuthally averaged features of the ring-like structure reported by 
Lada et al. (2005). The initial conditions for this simulation are a truncated Bonnor-Ebert 
sphere with boundary 
at radius $R_{_{\rm B}} = 3 a_{_0} / (4 \pi G \rho_{_{\rm C}})^{1/2}$, where $a_{_0}$ is the 
isothermal sound speed and $\rho_{_{\rm C}}$ is the central density. We set $a_{_0} = 
0.2\,{\rm km}\,{\rm s}^{-1}$ (corresponding to molecular gas at $T_{_0} = 10\,{\rm K}$) and 
$\rho_{_{\rm C}} = 0.2 \times 10^{-20}\,{\rm g}\,{\rm cm}^{-3}$ (corresponding to 
$n_{_{\rm C}} = 0.5 \times 10^3\,{\rm H}_2\,{\rm cm}^{-3}$). This gives a core radius 
$R_{_{\rm B}} = 0.4\,{\rm pc}$ and a total core mass $M_{_{\rm TOTAL}} =6.5\,{\rm M}_{_\odot}$. 
The cloud is in stable equilibrium with external pressure $P_{_{\rm EXT}} = P_{_0} = 
0.22 \times 10^{-12}\,{\rm erg}\,{\rm cm}^{-3} \equiv 0.16 \times 10^4\,k_{_{\rm B}}\,
{\rm cm}^{-3}\,{\rm K}$. The initial conditions are set up with a random but settled (i.e. 
non-crystalline) distribution of particles in detailed hydrostatic equilibrium, and the subsequent 
evolution is isothermal (i.e. $P = a_{_0}^2 \rho$, with uniform and constant $a_{_0}$). 
At $t = 0$, we increase $P_{_{\rm EXT}}$ at a constant rate, by a factor of 10, over a thirtieth 
of a crossing time, i.e. 
\begin{eqnarray}
P_{_{\rm EXT}} & = & \left\{ \begin{array}{lr}
P_{_0}\,, & t \leq 0\,; \\
\left( 1 \,+\, 270\,t / t_{_{\rm SC}}\right)\,P_{_0} \,,\;\; & 0 < t \leq t_{_{\rm SC}} / 30 \,; \\
10\,P_{_0} \,, & t > t_{_{\rm SC}} / 30 \,. \end{array} \right.
\end{eqnarray}
The sound-crossing time is given by $t_{_{\rm SC}} = R_{_{\rm B}} / a_{_0} = 2\,{\rm Myr}$, 
and so the new increased external pressure is reached after $\sim 0.07\,{\rm Myr}$.

We then explore how the evolution of the core is modified if it has a modest level of internal 
turbulence. This is motivated by the fact that the ring-like structure observed by Lada et al. presents 
 significant departures from an ideal axisymmetric ring. We therefore want to explore wheter turbulence
is able to produce similar patern and for which level of turbulence.
  
The turbulence is characterised by two parameters: (i) the ratio of turbulent energy, 
$E_{_{\rm TURB}}$, to self-gravitational potential energy, $\Omega$, i.e. $\alpha_{_{\rm TURB}} 
= E_{_{\rm TURB}}/|\Omega|$; and (ii) the exponent $n$ in the turbulent power spectrum, $P(k) 
\propto k^{-n}$. The turbulent parameters in the different simulations are given in Table 1
\begin{table}
\label{TAB:TURB}
\caption{Parameters of the initial turbulence in the different simulations. }
\begin{center}
\begin{tabular} {cccc}
Simulation & $\alpha_{_{\rm TURB}}$ & $n$ &  Figs. \\ \hline
 & & \\
1 & 0 & -  & 1\&2 \\
2 & 0.01 & 5/3  & 3 \\
3 & 0.02 & 5/3  & 4 \\
4 & 0.02 & 0    & 5 \\
\end{tabular}
\end{center}
\end{table}

\section{Simulation results}
\subsection{Spherical case}
Fig.~1 shows the profiles of volume-density, $n(r)$, inflow velocity, $u(r)$, 
and column-density, $N(b)$, generated by the non-turbulent simulation (Simulation 
1 in Table \ref{TAB:TURB}) at three times, $t = 1.170,\,1.183,\,{\rm and}\,1.198\,{\rm Myr}$ 
(i.e. well after the external pressure has reached its new increased value, $10P_{_0}$). Here 
$r$ is distance from the centre of the core and $b$ is the impact parameter of the line of sight
relative to the centre of the core. 
The density in the swept-up shell has been increased from $n \sim 0.3 \times 10^3\,{\rm cm}^{-3}$ 
to $n \sim 0.6 \times 10^4\,{\rm cm}^{-3}$, and the swept-up shell is travelling inwards at $u 
\sim 0.4\,{\rm km}\,{\rm s}^{-1}$ (i.e. Mach-2). 

If we adopt the relation $N = 0.2 \times 10^{22}\,{\rm cm}^{-2} \times A_{_{\rm V}}$ (where $N$ 
is the column-density of hydrogen nuclei in all forms, and $A_{_{\rm V}}$ is the visual extinction 
in magnitudes), we can convert the azimuthally averaged 
extinction profile obtained by Lada et al. (2004) from their near-IR measurements through G2 (their 
Fig. 3) into a column-density profile. The result of this conversion is shown as a dashed line 
in the righthand panels of Fig.~1. We see that the observed column-density profile 
is fitted best by the simulated column-density profile at $t = 1.183\,{\rm Myr}$. The 
simulated profile has a minimum of $N \sim 0.16 \times 10^{23}\,{\rm cm}^{-2}$ in the centre, 
and a maximum of $N \sim 0.24 \times 10^{23}\,{\rm cm}^{-2}$ at radii $r \sim 0.04\,{\rm pc}$, 
close to the maximum of the observed profile. We note that, with a spherically symmetric model, 
it requires a very deep -- approximately tenfold -- drop in the volume density $n$ at the centre 
to produce a $\sim 30\%$ dip in the column-density $N$; it is very unlikely that the subsonic 
velocities involved in an oscillating core could reproduce this feature.

It is  worth to stress that our model predicts a column density in the external part of 
the cloud ($x \ge 0.1$ pc) which is roughly two times lower than the observed column-density.
This discrepency could be due to the fact that we do not model the gas outside the core which may 
have, either a direct contribution to the total column-density or  could have been accreted 
in part by the core therefore enhancing the core density in its outer part. 

\begin{figure}
\label{FIG:SPECTRE}
\setlength{\unitlength}{1mm}
\begin{picture}   (40,65)
\includegraphics{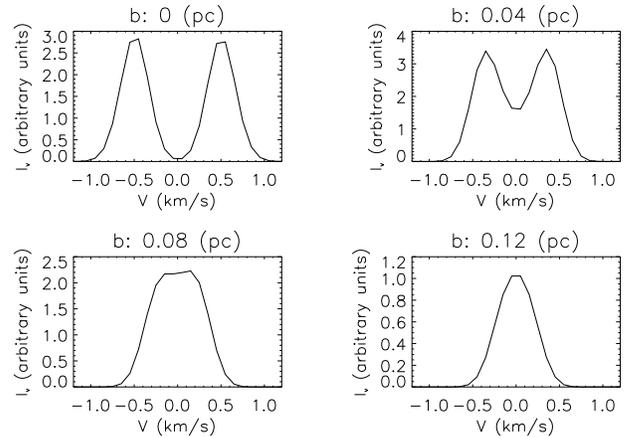}
  \end{picture}
\caption{$^{12}$C$^{18}$O ($J = 2 \rightarrow 1$) line profiles at time $t = 1.183\,
{\rm Myr}$, from lines of sight at different impact parameters $b$, relative to the 
centre of the simulated core. The impact parameters are $b = 0,\,0.04,\,0.08\,{\rm 
and}\,0.12\,{\rm pc}$.} 
\end{figure}

Lada et al. (2004) also report a single line profile for the $^{12}$C$^{18}$O ($J = 2 
\rightarrow 1$) transition from a line of sight which is displaced from the notional centre 
of symmetry of the core by $\Delta \delta = 68.5''$ and $\Delta \alpha = 69.''$, giving a
net angular displacement of $\sim 100''$. At a distance of $D = 150\,{\rm pc}$, this 
corresponds to an impact parameter $b \sim 0.08\,{\rm pc}$. We can therefore predict the 
line profile expected from the simulation, using
\begin{equation}
I_{_v} \;=\; \int j\left( n(r),T_{_0} \right)\,\left\{ \frac{dv'}{ds} \right\}^{-1}\,\phi(v'-v)\,dv' \,,
\end{equation}
where $v$ is the radial velocity (i.e. frequency) observed, $j\left( n(r),T_{_0} \right)$ 
is the integrated volume emissivity 
in the line, $s$ is distance along the line of sight (measured from the tangent point), $r(s) = 
\left( b^2 + s^2 \right)^{1/2}$, $v'(s) = u\left(r(s)\right) s / r(s)$ is the inflow velocity 
projected along the line of sight, and $\phi(v'-v)$ is the profile function (reflecting thermal, 
turbulent and natural broadening of the line).

Fig.~2 shows emission line profiles for $^{12}$C$^{18}$O ($J = 2 \rightarrow 1$) 
on lines of sight with impact parameters $b = 0,\,0.04,\,0.08\,{\rm and}\,0.12\,{\rm pc}$. The 
simulated line profile at $b = 0.08\,{\rm pc}$ matches the width of the broad line component 
observed by Lada et al. (2004) quite well. We concur with their interpretation of the narrow 
line component as due to a small-scale turbulent fluctuation, and we would not expect the present 
model to reproduce this. From the profiles presented in Fig.~2, it 
should be straightforward to test the model proposed here by measuring $^{12}$C$^{18}$O ($J = 2 
\rightarrow 1$) line profiles at different positions within the G2 core. Positions within $\sim 
50''$ of the centre should present double-peaked line profiles, due to the receding and 
approaching sides of the compression wave.


\setlength{\unitlength}{1cm}
\begin{figure}
\begin{picture}(0,16)
\put(0,8){\includegraphics[width=8cm]{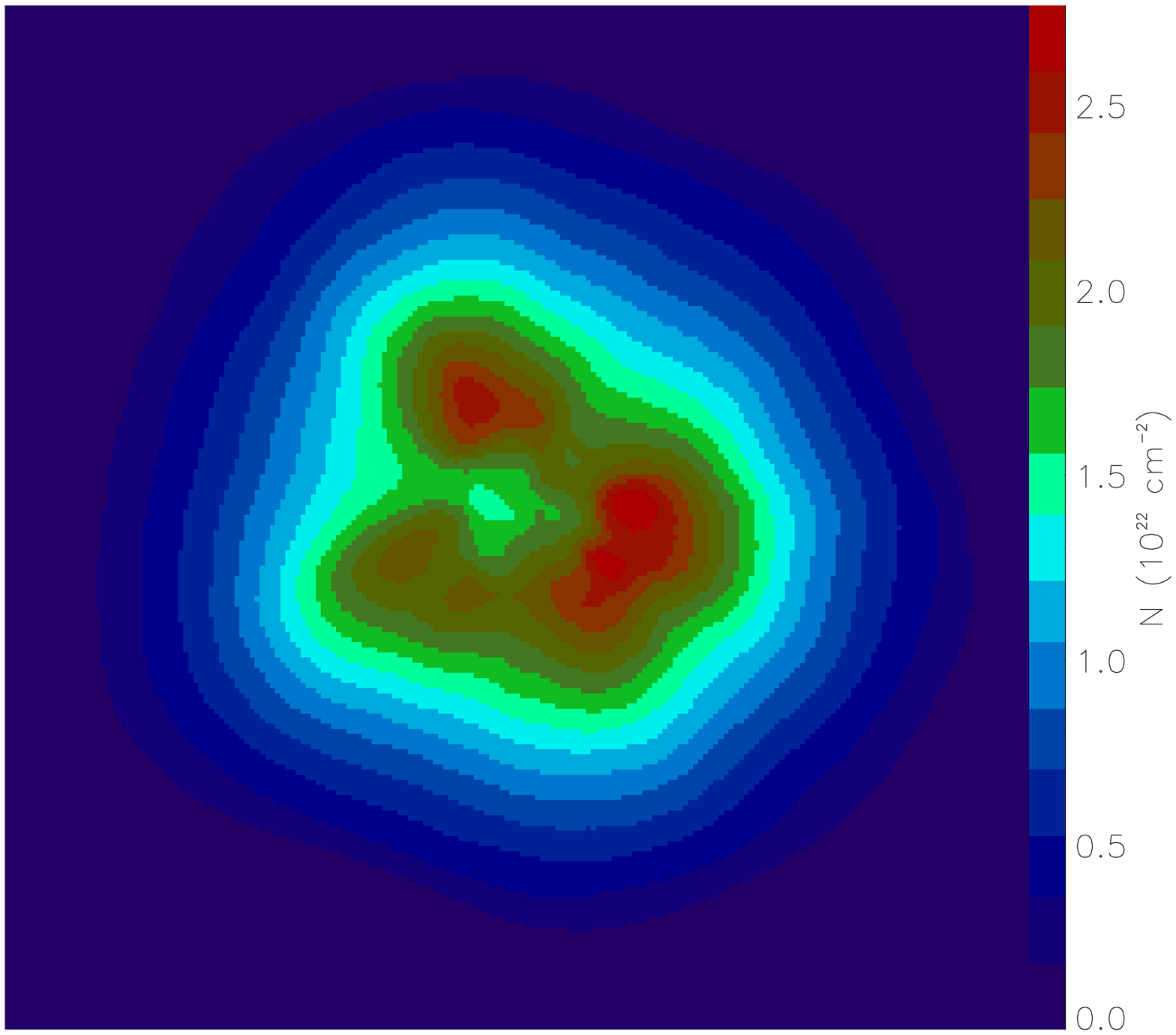}}
\put(0,0){\includegraphics[width=8cm]{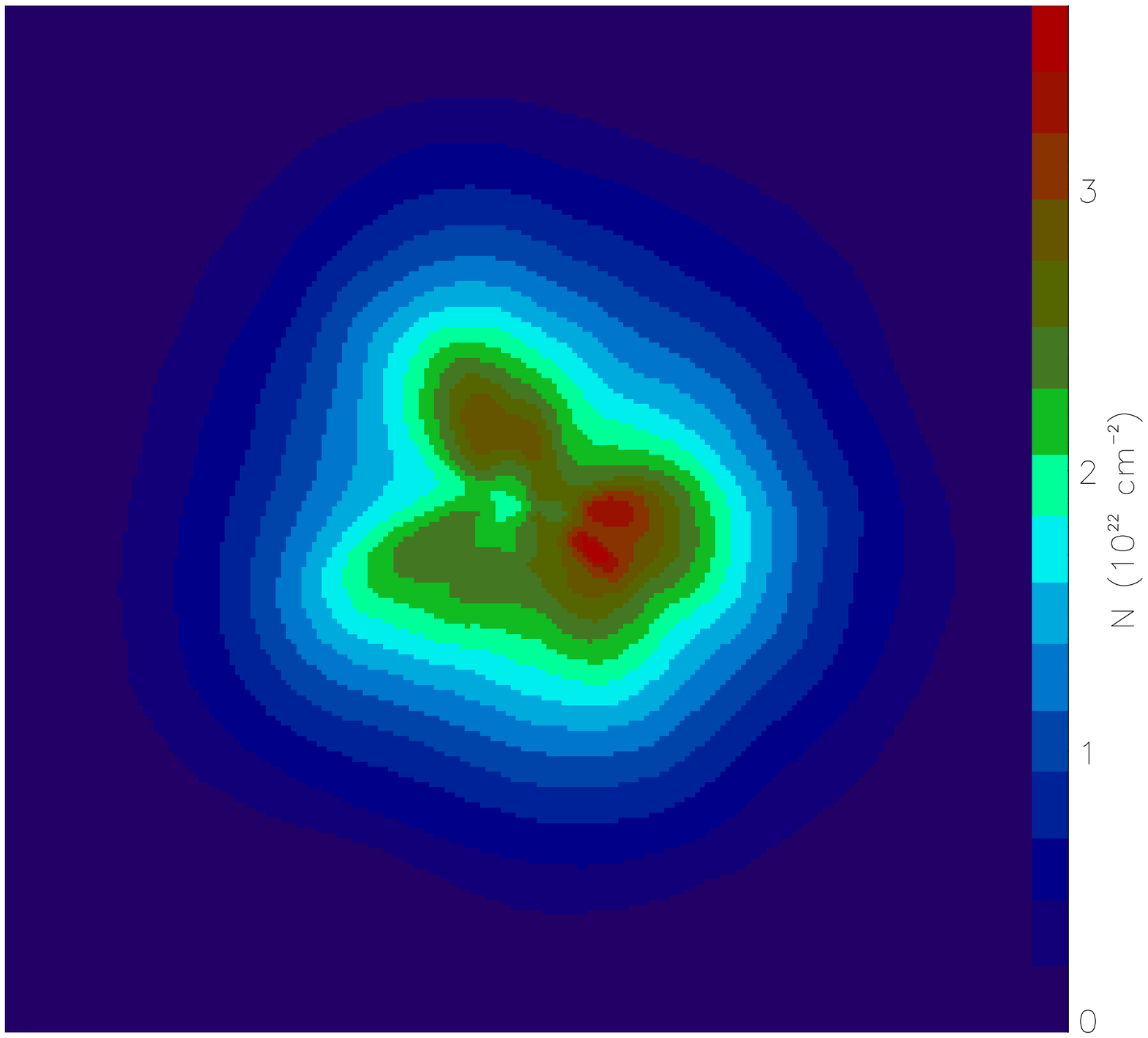}}
\end{picture}
\caption{Column density maps from simulations with initial turbulence.
 Parameters are: $\alpha_{_{\rm TURB}} = 0.01$, $n = 5/3$. 
Top panel is for time equal to $1.17\,{\rm Myr}$ whereas second panel is for
$1.18\,{\rm Myr}$. }
\label{turb01_5_3}
\end{figure}

\setlength{\unitlength}{1cm}
\begin{figure}
\begin{picture}(0,16)
\put(0,8){\includegraphics[width=8cm]{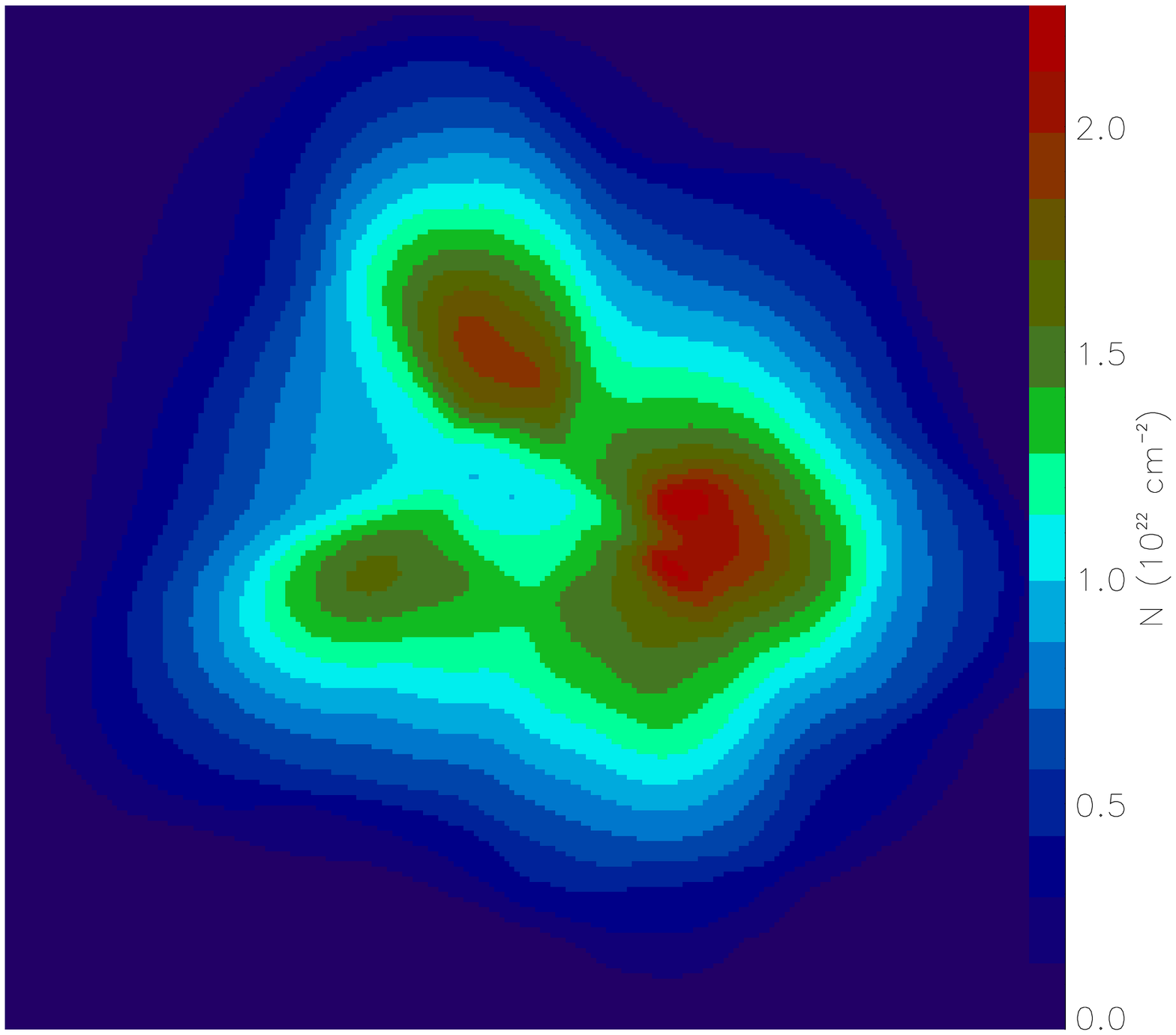}}
\put(0,0){\includegraphics[width=8cm]{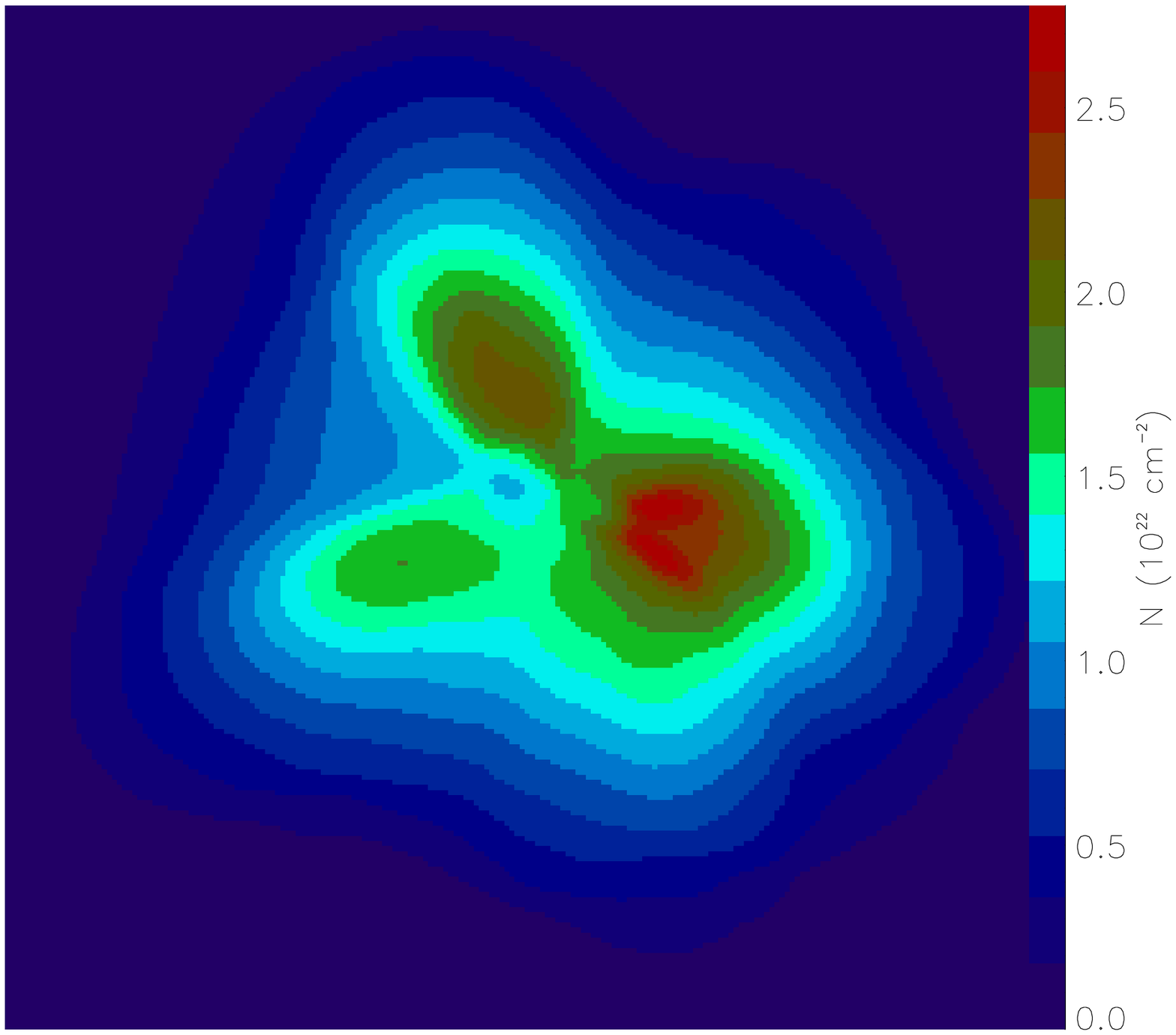}}
\end{picture}
\caption{Same as Fig.~\ref{turb01_5_3} for 
$\alpha_{_{\rm TURB}} = 0.02$, $n = 5/3$. 
Top panel is for time equal to $1.17\,{\rm Myr}$ whereas second panel is for
$1.185\,{\rm Myr}$. }
\label{turb02_5_3}
\end{figure}

\setlength{\unitlength}{1cm}
\begin{figure}
\begin{picture}(0,16)
\put(0,8){\includegraphics[width=8cm]{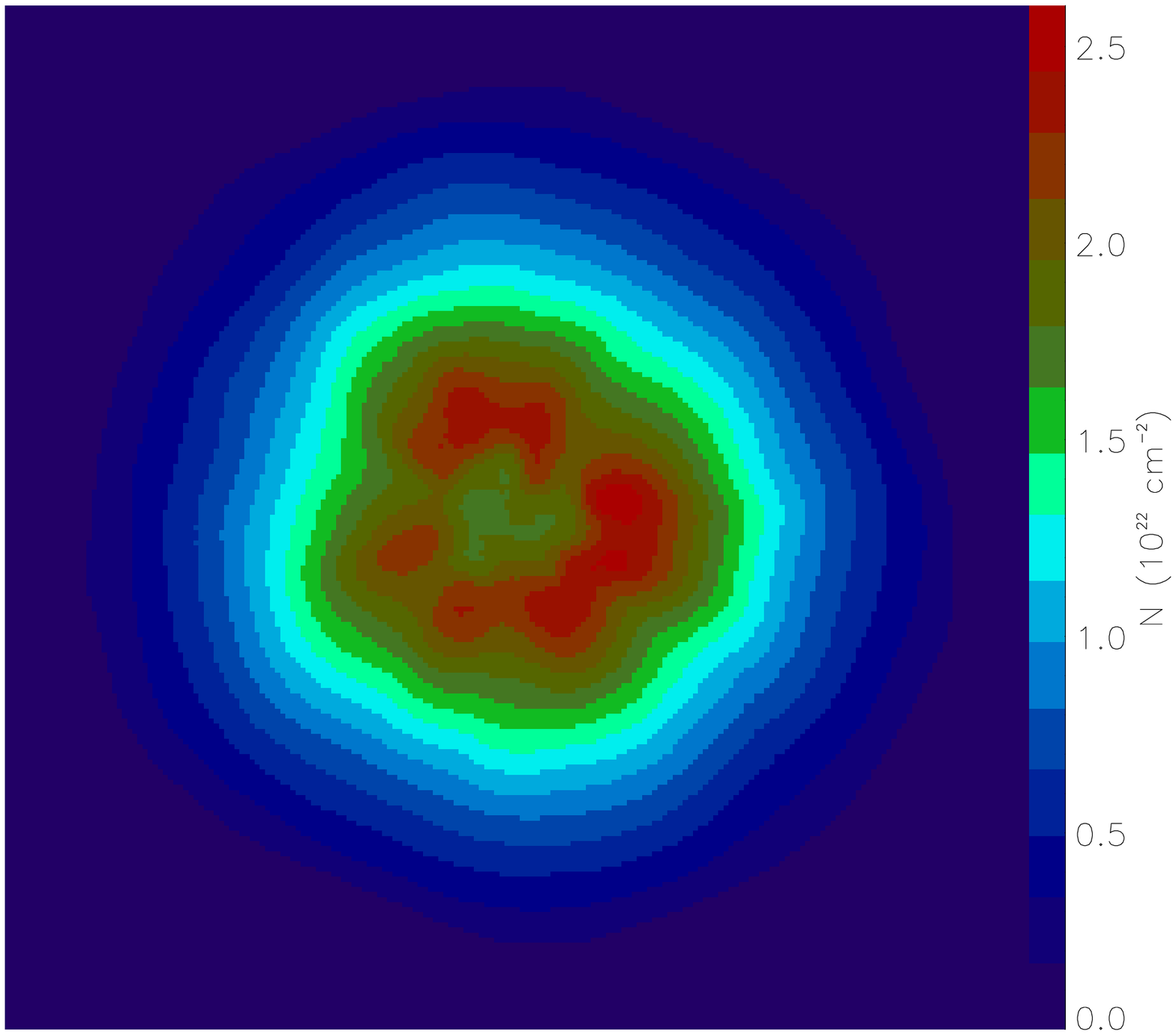}}
\put(0,0){\includegraphics[width=8cm]{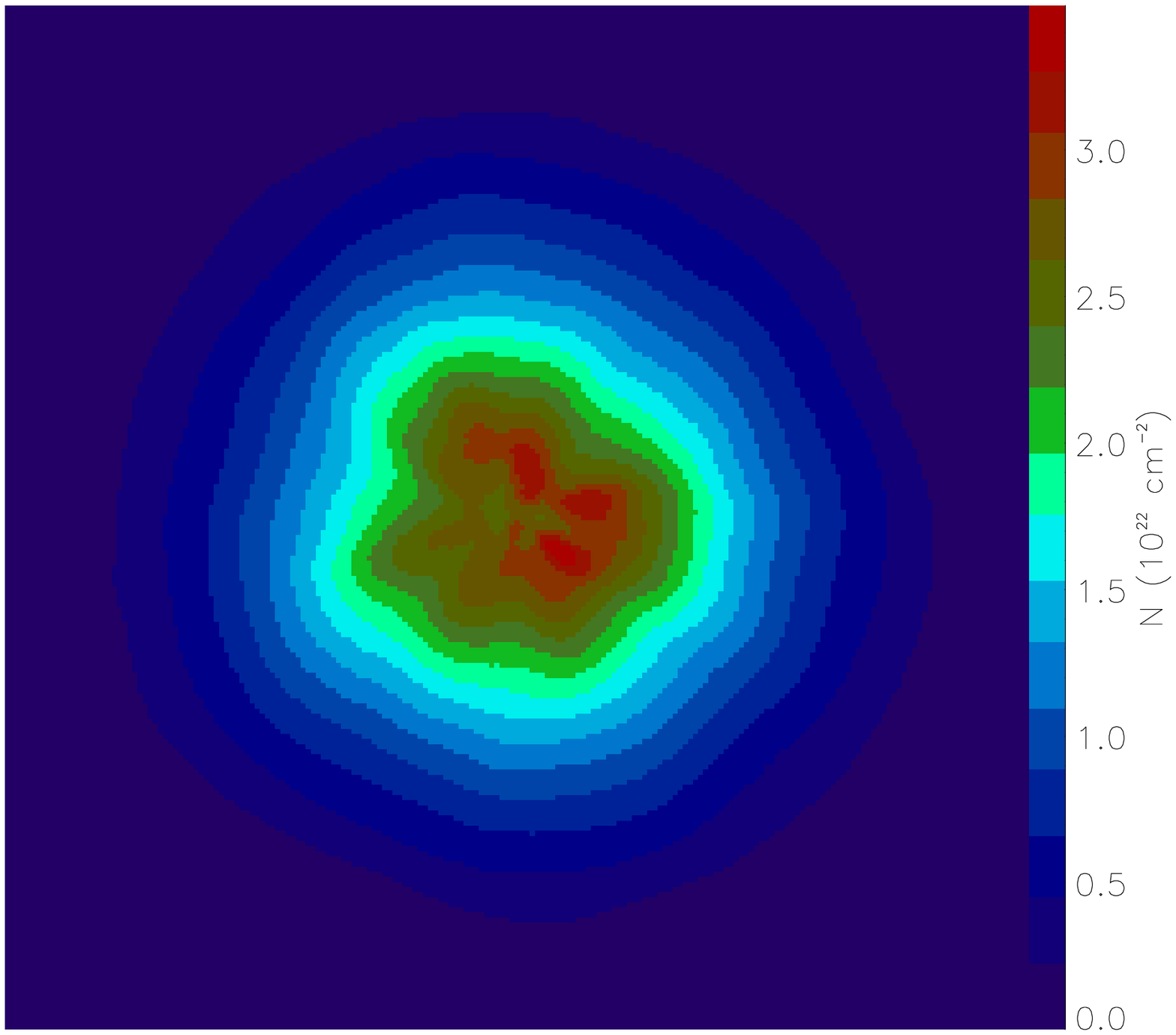}}
\end{picture}
\caption{Same as Fig.~\ref{turb01_5_3} for 
$\alpha_{_{\rm TURB}} = 0.02$, $n = 0$. 
Top panel is for time equal to $1.175\,{\rm Myr}$ whereas second panel is for
$1.20\,{\rm Myr}$.}
\label{turb02_0}
\end{figure}

\subsection{Effect of turbulence}
Here we present the results of three simulations in which the initial conditions 
are the same as in Simulation 1, except that a turbulent velocity field is superimposed at 
the outset. The figures all show column density maps. 
The different turbulence parameter 
combinations are given in Table 1. We note that the levels of turbulence 
invoked are very low $\alpha_{_{\rm TURB}} = 0.01\;{\rm and}\;0.02$. The turbulent 
power spectra have exponents $n = 5/3$ (corresponding to Kolmogorov) and $0$ (i.e. 
more power on small scales).

Fig.~3 displays the column densities at two times steps close to t=1.18 Myr which 
matches  best the observation. It shows that even with $\alpha_{_{\rm TURB}} = 0.01$ 
and $n=5/3$ (Simulation 2)  the ring is significantly distorted. It is still roundish but 
is fragmented in denser clumps. We note that this is actually the case for the ring 
observed by Lada et al.

Fig.~4 shows that with $\alpha_{_{\rm TURB}} 
= 0.02$ and $n=5/3$ (Simulation 3) the ring is lost rapidly and completely. 
The gas rapidly concentrates  in one single dense clump.

The only way to 
retrieve the ring with $\alpha_{_{\rm TURB}} = 0.02$
is to concentrate turbulent energy at short wavelengths by decreasing 
$n$. This is illustrated in Fig.~5, where $\alpha_{_{\rm TURB}} = 0.02$ and $n=0$ 
(Simulation 4), and the ring stays almost axisymmetric during a longer amount of time. 
However, such a flat power spectrum is unrealistic since usually Kolmogorov-like powerspectra
are found in observations and in numerical simulations. We note that for large values of 
$\alpha_{_{\rm TURB}}$ ($\ge 5 \%$), the ring is destroyed even with $n=0$.

We conclude that, within the scope of our model, the ring-like structure observed by 
Lada et al. (2005) requires the initial core to have a very low level of turbulence. 
This suggests that the initial core has been very settled before it was 
subjected to increased pressure so that there has been enough time for the 
turbulence to decay almost completely.

\section{Discussion and Conclusions}

We have modelled Tapia's Globule G2 in the Coalsack as a core which is responding to a rapid 
increase in external pressure. The resulting compression wave has not yet reached the centre, 
and therefore there is a central rarefaction which manifests itself as a dip in the extinction 
towards the centre, as observed by Lada et al. (2004). This model requires that the core was 
already in existence as a coherent entity with relatively low mean density, $\bar{n}_{_{\rm INIT}} 
\sim 0.4 \times 10^3\,{\rm cm}^{-3}$, and large extent, $r_{_{\rm INIT}} \sim 0.4\,{\rm pc}$, and 
that the external pressure acting on it increased by a factor $\sim 10$ (from $\sim 0.16 \times 
10^4\,k_{_{\rm B}}\,{\rm cm}^{-3}\,{\rm K}$ to $\sim 0.16 \times 10^5\,k_{_{\rm B}}\,{\rm cm}^{-3}
\,{\rm K}$) in a small fraction of a sound-crossing time, $t_{_{\rm SC}} / 30 \sim 0.07\,{\rm Myr}$. 
The advantages of such external triggers for synchronising the formation of stars in 
clusters and generating high multiplicity have been emphasised by Whitworth et al. (1996).

One possibility is that the Coalsack has been involved in, or even created by, a collision 
between two large-scale turbulent flows. To deliver the required large pressure increase in 
the required short time requires a flow with minimum density $n_{_{\rm MIN}} \sim 2\,{\rm 
cm}^{-3}$ and minimum speed $v_{_{\rm MIN}} \sim 6\,{\rm km}\,{\rm s}^{-1}$. Neither 
requirement is extreme. Indeed the models of Audit \& Hennebelle (2005) show that pressure 
increases of the magnitude invoked by our model are very common in the atomic gas from which 
molecular complexes like the Coalsack are presumed to form. 

A second possibility is that the core has fallen into a denser environment and been compressed by 
the increased ambient pressure. The local cloud within which G2 resides is Region II in the 
scheme of Nyman et al. (1989); see also Kato et al., (1999). This cloud has mass $M \sim 400\,
{\rm M}_{_\odot}$ and cross-sectional area $A \sim 9\,{\rm pc}^2$, so its mean surface density 
is $\bar{\Sigma} \sim M / A \sim 0.01 \,{\rm g}\,{\rm cm}^{-2}$, and therefore, assuming that 
it is self-gravitating, its central pressure must be $P_{_{\rm C}} \sim G \bar{\Sigma}^2 \sim 
0.5 \times 10^5\,k_{_{\rm B}}\,{\rm cm}^{-3}\,{\rm K}$. It follows that the core only needs to 
travel about one third of the cloud radius, i.e. $\sim 0.5\,{\rm pc}$, to experience the 
required increase in external pressure, and it can do this in $t_{_{\rm SC}} / 30 \sim 0.07\,
{\rm Myr}$ by travelling at $\sim 7\,{\rm km}\,{\rm s}^{-1}$, which is a typical bulk velocity 
for a cool cloud in the local interstellar medium. Since the measured radial velocity of G2 is 
$\sim - 6\,{\rm km}\,{\rm s}^{-1}$, and the measured radial velocity of the Coalsack is 
$\sim -\,5\;{\rm to}\;0\;{\rm km}\,{\rm s}^{-1}$ (Nyman et al. 1989), the relative radial 
velocity is in the range $\sim 1\;{\rm to}\;5\;{\rm km}\,{\rm s}^{-1}$. Thus if the net 
velocity of G2 relative to the Coalsack is $\sim 7\,{\rm km}\,{\rm s}^{-1}$, the chance of 
its radial component being in this range is $\sim 71\%$. Given that we have only one 
case, we cannot take statistical arguments any further, but this at least illustrates 
that the model is not based on a highly unlikely coincidence. We note that this estimation is based 
on the assumption that the main effect of the globule falling into a denser region would be 
the pressure increasement. However other processes like accretion of more material into the 
original globule could certainly have a significant impact and promote the formation of a core 
out of equilibrium.

We have also shown that within the scope of our model 
the initial level of turbulence in the core (before $t = 0$) must 
have been very low ($\alpha_{{\rm TURB}} \le 5 \%$) 
in order for the ring-like structure to develop in a coherent manner. 
Turbulence promotes the fragmentation of this ringlike structure seen in projection and 
generates dense clumps which look like some of the substructures observed by Lada et al. 
in their ringlike  structures. However we note that unlike in our simulations, 
 in the observations these overdensities seem to protrude from the ring.  We do not know at 
this stage the reason of this. This may be due to the setting up of turbulent fluctuations 
with random phases or to the incomplete treatment of the physics of the core since here we ignore
magnetic field and large scale anisotropy. Our numerical resolution could also be insufficient to 
describe accurately  such small scale structures.

We reiterate that we have not performed an extensive parameter search to obtain a best-fit model. 
The observational constraints (the azimuthally averaged, radial extinction profile; and a single, 
off-centre, approximately pencil-beam, line-profile) do not justify detailed modelling, so we are 
here only attempting a feasibility study. The next step in testing the model should be to perform 
further line observations to test the kinematic predictions of the model (see Fig.~2). We hope to 
obtain these observations soon. Another line of investigation which should be pursued is to explore 
numerically the departures from spherical symmetry which can be expected because the 
pressure increase acting on the core is likely to be anisotropic.

In the meantime, it is appropriate to contrast the predictions of our model with those of the 
magnetic model developed by Li \& Nakamura (2002). 

(i) The magnetic model requires that the core be viewed from close to the axis of symmetry. In 
contrast, our model can be viewed from any angle, since it is spherically symmetric.

(ii) The magnetic model requires the magnetic field to be mainly along the line of sight. Lada 
et al. (2004) show that the field component in the plane of the sky (estimated from the dispersion 
in polarisation angles of background stars, using the Chandraskhar-Fermi equation) is rather close 
to the equipartition value, and they point out that this rather large field component in the plane 
of the skyis hard to reconcile with the idea that the main component is along the line of sight.

(iii) G2 is located towards the edge of the cloud complex, so it is difficult to see how it can 
evolve quasistatically in the ordered way described by Li \& Nakamura (2002). This presumably 
requires the field to be anchored in some larger-scale structure, but no such structure is evident 
on the CO maps of Nyman et al. (1989) and Kato et al. (1999).

 The model proposed here falls naturally within the category `spherical geometry: a core 
in transition' defined by Lada e al. (2004). The question then arises as to whether the core 
will eventually settle into an equilibrium state (a Bonnor-Ebert sphere) or whether it will 
continue contracting to form a star (or stars) at the centre. 
Indeed as in Hennebelle et al. (2003, 2004) later times of the simulations show  that when the 
compression wave converges on the centre of the core it will create a primary protostar. The compression wave is expected to reach the centre in about 
$40,000\,{\rm years}$.


\begin{acknowledgements}

We thank Maryvonne Gerin for drawing the Lada et al. paper to our attention, and Charlie Lada 
for useful discussions. This paper was written while APW was a visitor at the \'Ecole Normale 
Sup\'erieure, and he is very grateful for the support and hospitality he received during his visit.
SPG is supported by a UKAFF Fellowship. We thank Ian Bonnell, the referee, for his comments which have
significantly improved the original manuscript.
\end{acknowledgements}

\end{document}